\documentclass[preprint]{elsarticle}
\usepackage{amssymb}
\usepackage{graphicx}
\usepackage{tikz}
\usepackage{sistyle}
\usepackage{color,colortbl}
\usepackage{caption}
\usepackage{subcaption}
\usepackage{ulem,cancel}
\usepackage[pdf]{pstricks}
\usepackage{epsfig,epsf,amsfonts,amscd}

\newcommand{\Z}{{\mathbb{Z}}} 

\usepackage{amstext}

\biboptions{square,comma}

\begin{document}

\begin{frontmatter}

\title{Spectral properties of the ladder-like Josephson junction array}
\author[db]{Daryna Bukatova}
\ead{daryna.bukatova@gmail.com}

\author[is]{Ivan O. Starodub}
\ead{starodub@bitp.kyiv.ua}

\author[yz]{Yaroslav Zolotaryuk\corref{cor1}}
\cortext[cor1]{Corresponding author}
\ead{yzolo@bitp.kyiv.ua}

\address
{Bogolyubov Institute for Theoretical Physics, National Academy of
Sciences of Ukraine, Kyiv 03143, Ukraine}
\begin{abstract}
In this paper theoretical analysis of the ladder-like multirow array of
inductively coupled Josephson junctions is presented. An external dc 
current
is applied at the top to each of the columns of the array 
and is extracted at the bottom of that column. 
The density of states of the Josephson plasma waves has a $\delta$-function term due to the flat band and $3N-2$ singularities where
$N$ is the number of rows.
The spatial distribution of the amplitudes 
of the plasmon wave is computed analytically for
any given value of the wavenumber $q$. It is expressed through the
orthogonal polynomials that are similar but not identical to the
Chebyshev polynomials.
\end{abstract}


\end{frontmatter}
\section{Introduction}
\label{intro}

Systems with dispersionless or flat bands have been an important
research topic in recent years \cite{laf18aip}. After the initial
theoretical predictions for the fermionic systems 
\cite{s86prb,l89prl,ks90jetpl}, research in this area has spred further
into Josephson junctions arrays \cite{phbgpp08prb,af19jpa}, Dirac
materials \cite{hv11jetp,v18jetpl,ggo21prb}, photonic lattices 
\cite{vcm-irm-cwsm15prl}, magnets \cite{dhr07prb} and other materials. 
In many cases the flat bands have been observed experimentally.

Ladders of inductively coupled Josephson junctions also support
flat bands. They have been studied in connection with various 
nonlinear wave phenomena such as
fluxons \cite{yls93prb,ldhhbb94prb}, discrete breathers
\cite{fmmfa96epl,tmo00prl,baufz00prl,mffzp01pre} 
and meandering \cite{acfflu99prl}. It has been shown 
first in  \cite{fs99jpc,baufz00prl,mffzp01pre} that a completely flat band
can appear in the simplest JJ ladder with two rows. 
In \cite{bz22jpc} the 
dispersion law for the Josephson plasma waves in the 
general case of $N$ rows that has a $N$-fold degenerate flat band 
was obtained. 
We remark that besides the weak superconductivity, the
 ladder systems have been researched actively 
in various areas of modern physics ranging from spin
ladders \cite{d99rpp,smkuf22natcommp} to integrable systems
\cite{v21pla,vv24ujp}. 

This work is focused on more deep investigation of the linear
wave properties in the multi-row ladder-like Josephson junction arrays.
We derive the plasmon density of states and obtain a general analytical
formula for the plasma wave amplitude. We also compute the
complete set of the plasmon amplitudes in the horizontal and
vertical subsystems.

The paper is organized as follows. The model of the ladder-like Josephson
junction array  
is presented in the next section. Section \ref{sec3} is devoted
to the presentation of the main results. Discussions and conclusions are
presented in the last section.

\section{Model and equations of motion}
\label{model}

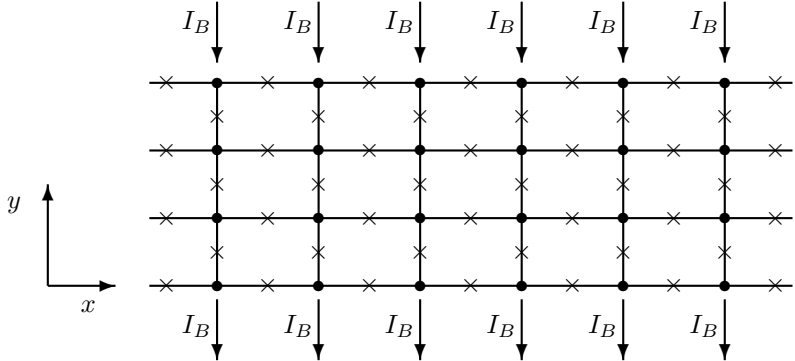
\begin{figure}[htb]
\setlength{\unitlength}{9.cm}
\begin{picture}(0.7,0.53)
\thicklines
\multiput(0.2, 0.0)(0,0.1){4}{
\multiput(0.1, 0.1)(0.15,0){6}{\circle*{0.015}}}
\multiput(0.2, 0.1)(0,0.1){4}{\line(1, 0){0.950}}
\multiput(0.3, 0.1)(0.15,0){6}{\line(0, 1){0.3}} 
\multiput(0,0)(0,0.1){4}{
\multiput(0.225, 0.1)(0.15,0){7}{\makebox(0,0){$\times$}}}
\multiput(0,0)(0.15,0){6}{
\multiput(0.3, 0.15)(0.0,0.1){3}{\makebox(0,0){$\times$}}}
\multiput(0.3,0.08)(0.15,0){6}{\vector(0,-1){0.09}}
\multiput(0.3,0.52)(0.15,0){6}{\vector(0,-1){0.09}}
\multiput(0.27, 0.04)(0.15,0){6}{\makebox(0,0){$I_B$}}
\multiput(0.27, 0.49)(0.15,0){6}{\makebox(0,0){$I_B$}}
\put(0.05,0.1){\vector(1,0){0.1}}
\put(0.05,0.1){\vector(0,1){0.15}}
\put(0.11,0.07){\makebox(0,0){$x$}}
\put(0.0,0.22){\makebox(0,0){$y$}}
\vspace{0.7cm}
\end{picture}
\caption{Schematic representation of the biased ladder-like Josephson
junction array with $N=4$ rows. The junctions are marked by $\times$.}
\label{fig1}
\end{figure}

A ladder-like quasi-one-dimensional 
Josephson junction array (JJA) or a Josephson transmission line is considered.
The array forms a rectangular-shaped network as shown in Fig. \ref{fig1},
where the junctions (denoted by $\times$) 
lie the links that connect the vertices of the lattice.
It is assumed that the array has a finite number of rows in the
$Y$ direction. We denote this number by $N$. The number of columns
in the $X$ direction is assumed to be very large, $L_x\gg N$.
For the sake of simplicity we will consider the $L_x\to \infty$ limit.
Each junction that belongs to the $k$th row and $m$th column 
is described by the Josephson phase $\phi^{(v,h)}_{m,k}$, 
which is the difference of the wavefunction phases of the superconductors
that form the particular junction.
The letters in the superscript correspond to the {\it vertical} (v)
or {\it horizontal} (h) junctions. 
Each column is biased by the current $I_B$.
The array is anisotropic in the sense that the junctions that
belong to the rows (horizontal junctions) and to the columns (vertical junctions)
have different properties. This anisotropy is characterized by the
parameter $\eta$:
\begin{equation}\label{1}
\eta\equiv \frac{I^{(h)}_c}{I^{(v)}_c} = \frac{C_h}{C_v},
\end{equation}
where $I^{(h,v)}_c$ are respective critical currents and $C_{h,v}$
are respective capacitances of the horizontal and vertical junctions.
%
%
%
The two important limiting cases of the anisotropy parameter are:
(i) $\eta\to 0$ - Josephson phases oscillate mainly in the 
horizontal subsystem and  (ii) $\eta \to \infty$ - oscillations of
the vertical junctions dominate over the horizontal ones.

The temporal dynamics of each Josephson junctions in the arrays
can be described by the resistively and capacitatively shunted 
junction model \cite{barone82}. Within this model this dynamics is governed
by the set of differential equations
\begin{equation}\label{2a}
\frac{C_{v,h}\hbar}{2eI_c^{(v,h)}}\ddot{\phi}^{(v,h)}_{m,k} +
\sin \phi^{(v,h)}_{m,k}=\frac{I^{(v,h)}_{m,k}}{I^{(v,h)}_c},
\end{equation}
where $I^{(v,h)}_{m,k}$ is the current passing through the 
$\phi^{(v,h)}_{m,k}$ junction. The dissipative term 
$\hbar \dot{\phi}^{(v,h)}_{m,k}/(2eR_{v,h})$ is omitted because
it will only add the imaginary decay factor to the dispesion
laws that we are goint to study.
Together with the flux quantization rule
inside the array loop and the Kirchoff laws, a set of coupled 
discrete sine-Gordon equation is derived. The derivation procedure
is based on the paper \cite{bw98prb}. The set of equations is written
explicitly in \cite{bz22jpc}, it is quite cumbersome, so we do not
repeat it here. 
As shown in Fig. \ref{fig1}, the constant bias current $I_B$ is injected
on the top of each column and, consequently, is extracted at the bottom of
that column. Similarly to \cite{bz22jpc} we work under the assumption of 
the uniform stationary current flow along each column. An
elementary cell of the JJA consists of $2N-1$ junctions: $N$ horizontal
and $N-1$ vertical.

It is convenient
to introduce the dimensionless time $\tau=t\omega_p$, where $\omega_p$
is the Josephson plasma frequency. We also introduce
the dimensionless inductance parameter that measures the array discreteness
in $X$ direction. Both these parameters are written as follows:
\begin{equation}\label{6}
 \omega_p =\sqrt{\frac{2 \pi I_c^v}{\Phi_0 C^v}},
 \;\;\;\beta_L= \frac{2\pi L I^{(v)}_c}{\Phi_0},\;\; \Phi_0=\frac{\pi\hbar}{e},
\end{equation}
where $\Phi_0$ is the magnetic flux quantum.
The dispersion 
law is obtained after linearization of the equations of motion
around the ground state
\begin{equation}\label{6a}
\phi^{(v)}_{m,k}=\arcsin \left (\frac{I_B}{I^{(v)}_c}\right)\equiv
\arcsin\gamma, k \in \overline{1,N-1};
\;\;\; \phi^{(h)}_{n,k}=0, \,k \in \overline{1,N}; \; m \in \Z.
\end{equation}
This ground state describes the situation when the external bias flows
uniformy along each of the columns.

\section{Properties of the Josephson plasma waves}
\label{sec3}

In this Section we discuss various properties of Josephson plasma waves
(plasmons) in the quasi-one-dimensional ladder-like array.

\subsection{Plasmon spectrum and its main properties}

Josephson plasmons are elecromagnetic waves that propagate
along the array in the $X$ direction.
At this point we remind the Josephson plasmon dispersion law for the $N$-row
JJA obtained in \cite{bz22jpc}.
This dispesion law consists is derived from the discrete system of the
sine-Gordon equations that are 
linearized around the groundstate (\ref{6a}). It consists of $2N-1$ branches:
\begin{eqnarray}
\label{7}
&&\omega_0=1,\\ 
\label{8}
&&\omega^2_{\pm n}(q)=W^+_n(q)\pm
 \sqrt{{W^-_n(q)}^2+2(1+\alpha_{n+1})\frac{1-\cos q}{\eta\beta_L^2}},\\
&& W^\pm_n(q)=\frac{1}{2}\left [1\pm \left(\omega^2_i(q)+\frac{1+\alpha_{n+1}}{\eta\beta_L} \right) \right],\\
&&\alpha_n=1-2\cos \frac{\pi(n-1)}{N} ,\; n=\overline{1,N-1} .
\end{eqnarray}
In Eq. (\ref{8}) the sign of the superscript in $\omega_{\pm}$ is
defined only by the sign near the radical and not by the
superscripts in $W^\pm$.
The function $\omega_i(q)$ is the plasmon dispersion law of the simple dc biased
parallel JJA with only one row of vertical junctions \cite{wzso96pd}:
\begin{equation}\label{8a}
\omega_i(q)=\sqrt{\sqrt{1-\gamma^2}+\frac{2}{\beta_L}(1-\cos q)}.
\end{equation}
The coefficients $\alpha_n$ always lie in the interval $]-1,3[$. For
$N=2$ rows there is only one coefficient $\alpha_2=1$, for $N=3$
there are two coefficients, $\alpha_2=0$, $\alpha_3=2$, for $N=4$ 
$\alpha_n=1-2\cos\frac{\pi(n-1)}{4}\Rightarrow$ $\alpha_2=1-\sqrt{2}$,
 $\alpha_3=1$, $\alpha_4=1+\sqrt{2}$ and so on.

The Josephson plasmon spectrum consists of $2N-1$
branches. Within the enumeration used in Eqs. (\ref{7})-(\ref{9}) these
branches are positioned as follows:
(i) {\it highly dispersive} branches $1<\omega_1<\omega_2<\cdots \omega_{N-1}$;
(ii) flat band with the Jopsehson plasma frequency $\omega_0=1$; 
(iii) {\it almost flat} branches $\omega_{-N+1}<\cdots<\omega_{-2}<\omega_{-1}<1$.
Thus, the highly dispersive branches are numbered by the positive
superscripts, lie above the flat band and look similar to the dispersion law of the simple parallel array $\omega_i(q)$ but are shifted  
from each other. In the unbiased case $\gamma=0$ all
the $\omega_{-|n|}$ branches become flat:
$\omega_{-N+1}(q)=\omega_{-N+2}(q)=\cdots=\omega_{-2}(q)=\omega_{-1}(q)=
\omega_0=1$. Otherwise they form a very narrow subband
that is confined in the following frequency intherval:
$(1-\gamma^2)^{1/4}\le \omega_{-n} <\omega_{-n}(\pi)<1$.
Even for rather high values of $\gamma$ this interval will be much
smaller than $1$.

\begin{figure}[htb]
\includegraphics[scale=0.28]{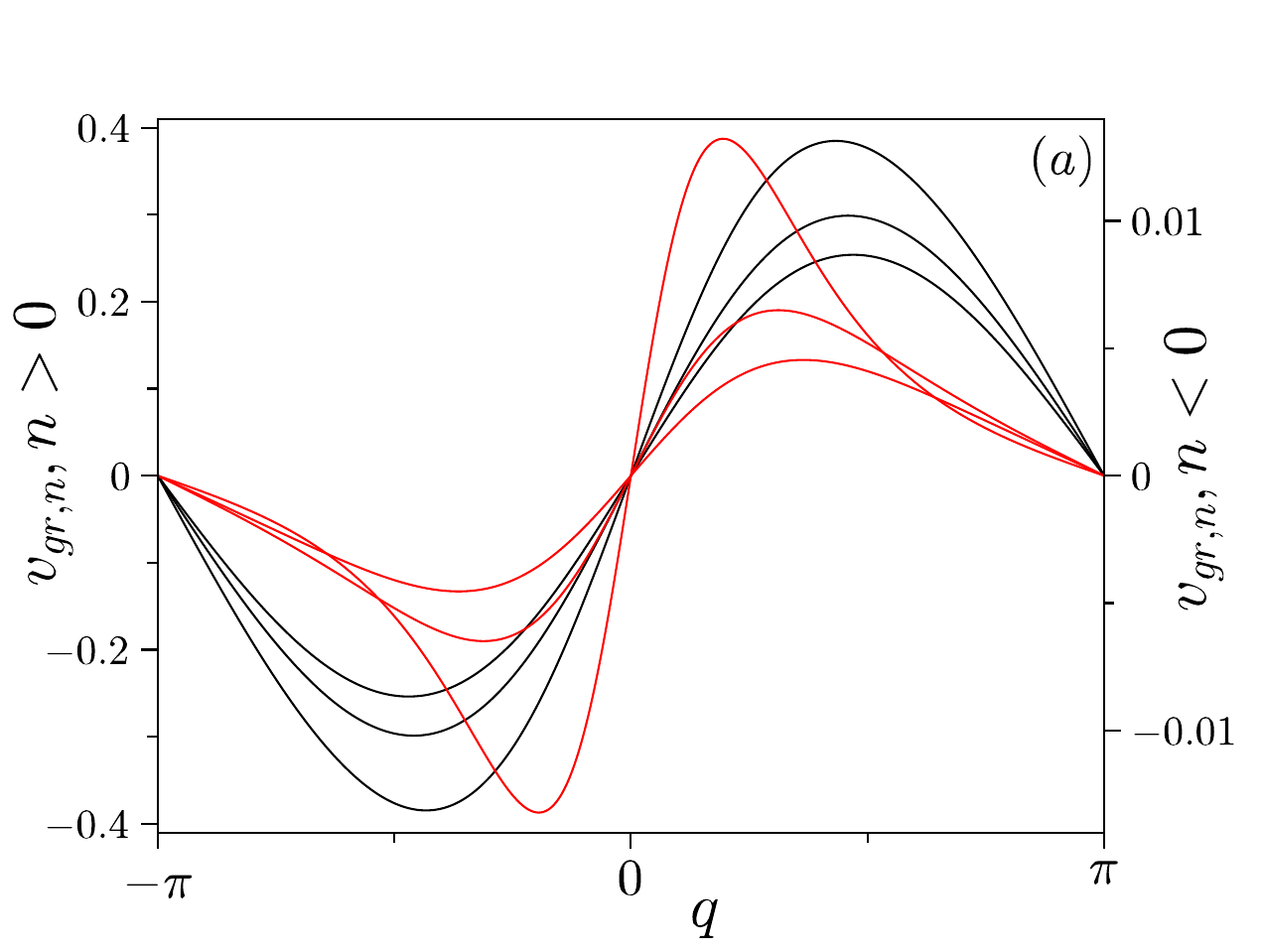}
\includegraphics[scale=0.28]{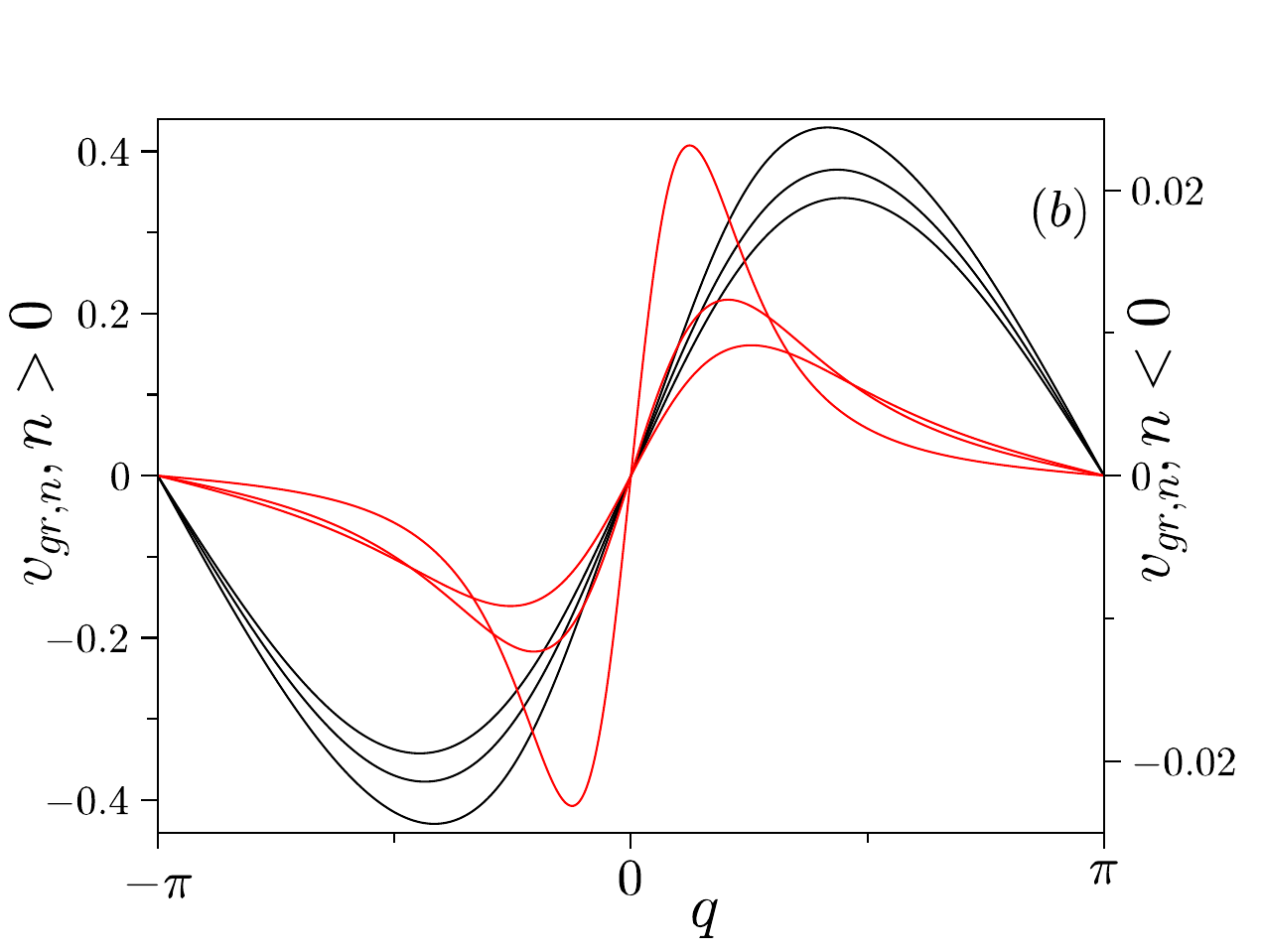}\\
\begin{center}
\includegraphics[scale=0.28]{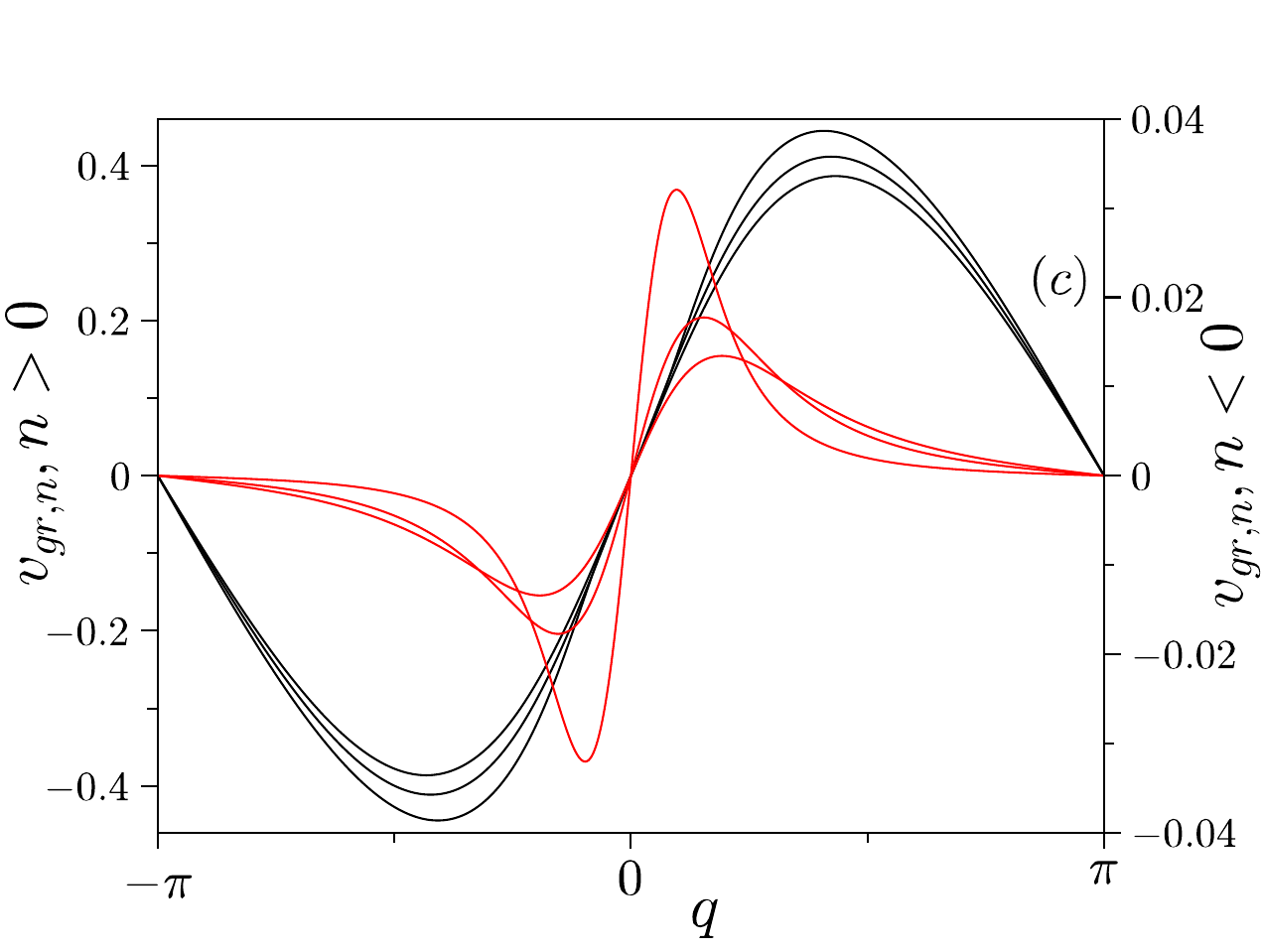}
\end{center}
\caption{The group velocities for the strongly dispersive branches,
$n=1,2,3$ (black, left scale) and almost flat branches 
$n=-3,-2,-1$ (red,right scale)
for $\gamma=0.3$, $\beta_L=1.5$, $N=4$ and $\eta=0.5$ (a), $\eta=1.5$
(b) and $\eta=3$ (c).}
\label{fig2}
\end{figure}

Group velocity characterizes the speed of the plasmon wavepackage
propagation. By looking at it one can have an idea how different
Josephson plasmon modes transport energy. The group velocity for
each of the $\omega_n(q)$ modes of (\ref{7})-(\ref{8}) has the
following form:
\begin{eqnarray}
&&v_{gr,0}(q)=0, \\
\nonumber
&&v_{gr,\pm n}(q)=\frac{d\omega_{\pm n}(q)}{dq}=
\frac{\sin q}{2\beta_L\omega_{\pm n}(q)}\left [1\pm \right. \\
&& \left. \pm
\frac{\omega^2_i(q)+\frac{\alpha_{n+1}}{\eta\beta_L}-1}{2\sqrt{{W^-_n(q)}^2+
\frac{2(1+\alpha_{n+1})}{\eta\beta_L^2}(1-\cos q)}} \right],\; n=\overline{1,N-1}.
\end{eqnarray}
It is obviously zero for the flat branch $\omega_0=1$, similarly
$v_{gr,-n}=0$ for the almost flat branches if the array is
not biased, $\gamma=0$. If $\gamma\neq 0$, 
the respective velocities for  the $n\neq 0$ modes are shown in Fig. \ref{fig2}.
There are three curves for each of the subbands, for the 
strongly dispersive branches $\omega_{1,2,3}$ and for the almost flat
branches $\omega_{-1,-2,-3}$. 
%
%
Since $\omega_1<\omega_2<\omega_3$, the same is true for the
group velocities: $v_{gr,1}<v_{gr,2}<v_{gr,3}$.
The group velocity for the strongly dispersive modes, $v_{gr,|n|}$,
does not deviate strongly from the $\sin q$ dependence. The maximal
values of $v_{gr,|n|}(q)$ are positioned near $q=\pi/2$. The situation is more
complex for the almost flat modes. The dispersion law for these
modes is parabolic near the origin $q=0$ but then reaches a plateau.
The inflexion point of the $\omega_{-|n|}(q)$ dispersion laws which is
the maximum of the group velocity shifts closer to the origin as 
$\eta$ increases. It should be noted that the 
following is true for the almost flat modes only close to the origin:
$v_{gr,-1}>v_{gr,-2}>v_{gr,-3}$. Away from the maximum these velocities
begin to cross.
The group velocity for the 
highest mode $\omega_{-1}$ is the sharpest and its maximum is the closest to the
origin. Thus, we conclude that the group velocity for the
almost flat modes is at least one order of magnitude smaller than for the
strongly dispersive modes. However, the almost flat modes, especially
the mode $\omega_{-1}$, exhibit a sharp peak not far from the
 $q=0$ point.

\subsection{Plasmon density of states}

The density of states (DoS) in various condensed matter systems 
contains useful information that can be accessed experimentally. 
We compute the Josephson plasmon density of states (DoS) per unit cell 
as a function of the frequency 
from the dispersion laws (\ref{7})-(\ref{8}).
The general expression for the plasmon DoS reads
\begin{equation}\label{dos0}
\rho(\omega)=\frac{1}{\pi}\left \{\sum_{n=-N+1;n\neq 0}^{N-1}
\frac{\theta[\omega-\omega_{n}(0)]-\theta[\omega-\omega_{n}(\pi)]}{\left|\frac{d\omega_{n}(q)}{dq}\right|}
+\delta(\omega-1) \right\}\,.
\end{equation}
The sum runs over the all non-flat bands $n\neq 0$, and the 
flat branch contribution is represented by the Diracs $\delta$-function
term $\delta(\omega-1)$. The Heaviside $\theta$-function is used to
account for the overlaping branches explicitly. This point will be 
discussed below.

Before discussing the plasmon DoS for the general array with an
arbitrary number of rows let us mention the simplest case of the 
JJ array of parallelly shunted junctions. The dispersion law
for the plasmons, $\omega_i(q)$ is given by equation (\ref{8a}). 
The respective density of states reads
\begin{equation}\label{dos00}
\rho(\omega)=\frac{\beta_L\omega}{\pi\sqrt{1-\left[1-\frac{\beta_L}{2}(\omega^2-\sqrt{1-\gamma^2}) \right]^2}}.
\end{equation}
It it not hard to notice that this function has two singularities
at the edges of the plasmonic band: $\omega_i(0)=(1-\gamma)^{1/4}$
and $\omega_i(\pi)=\sqrt{\sqrt{1-\gamma^2}+4/\beta_L}$. There is
a minimum somewhere between these two singularities.

By extracting from the dispersion law (\ref{8}) the dependence of the wavenumber $q$ on the frequency
$\omega$ it is possible to write down an exact expression for the 
plasmon density of states:
\begin{equation}\label{13}
\rho(\omega)=\frac{1}{\pi}\left\{2\beta_L\sum_{n=-N+1;n\neq 0}^{N-1}
\frac{\{\theta[\omega-\omega_{n}(0)]-\theta[\omega-\omega_{n}(\pi)]\}}
{[1+\mbox{sign}(n)~ Z_n(\omega)]S_n(\omega)}\omega
+\delta(\omega-1)\right\},
\end{equation}
where the auxilliary functions are given by
\begin{eqnarray}
&&Z_n(\omega)=\frac{\sqrt{1-\gamma^2}+\omega_{|n|}^2(0)-2+\bar{C}_n(\omega)}{\sqrt{\left[\sqrt{1-\gamma^2}-\omega_{|n|}^2(0)+\bar{C}_n(\omega)\right]^2+4\left[\omega_{|n|}^2(0)-1\right ]\bar{C}_n(\omega)}},\\
&&S_n(\omega)=\sqrt{1-\left[1-\frac{\beta_L}{2} \bar{C}_n(\omega)\right]^2},
\;\; \omega_{|n|}^2(0)=1+\frac{1+\alpha_{|n|+1}}{\eta \beta_L}, 
\label{15} \\
&&\bar{C}_n(\omega)=\frac{\left[\omega^2-\omega_{|n|}^2(0)\right]\left(\omega^2-\sqrt{1-\gamma^2}\right)}{\omega^2-1},\; n=\pm 1,\ldots,
\pm (N-1).
\end{eqnarray}
and the $\mbox{sign}(n)$ is used in order to take into account
 the $n>0$ or $n<0$ branches. The quantity $\omega_{|n|}^2(0)=[1+(1+\alpha_{|n|+1})/(\eta\beta_L)]$ is the minimum frequecy for the
 $n$th strongly dispersive branch.

\begin{figure}[htb]
\includegraphics[scale=0.5]{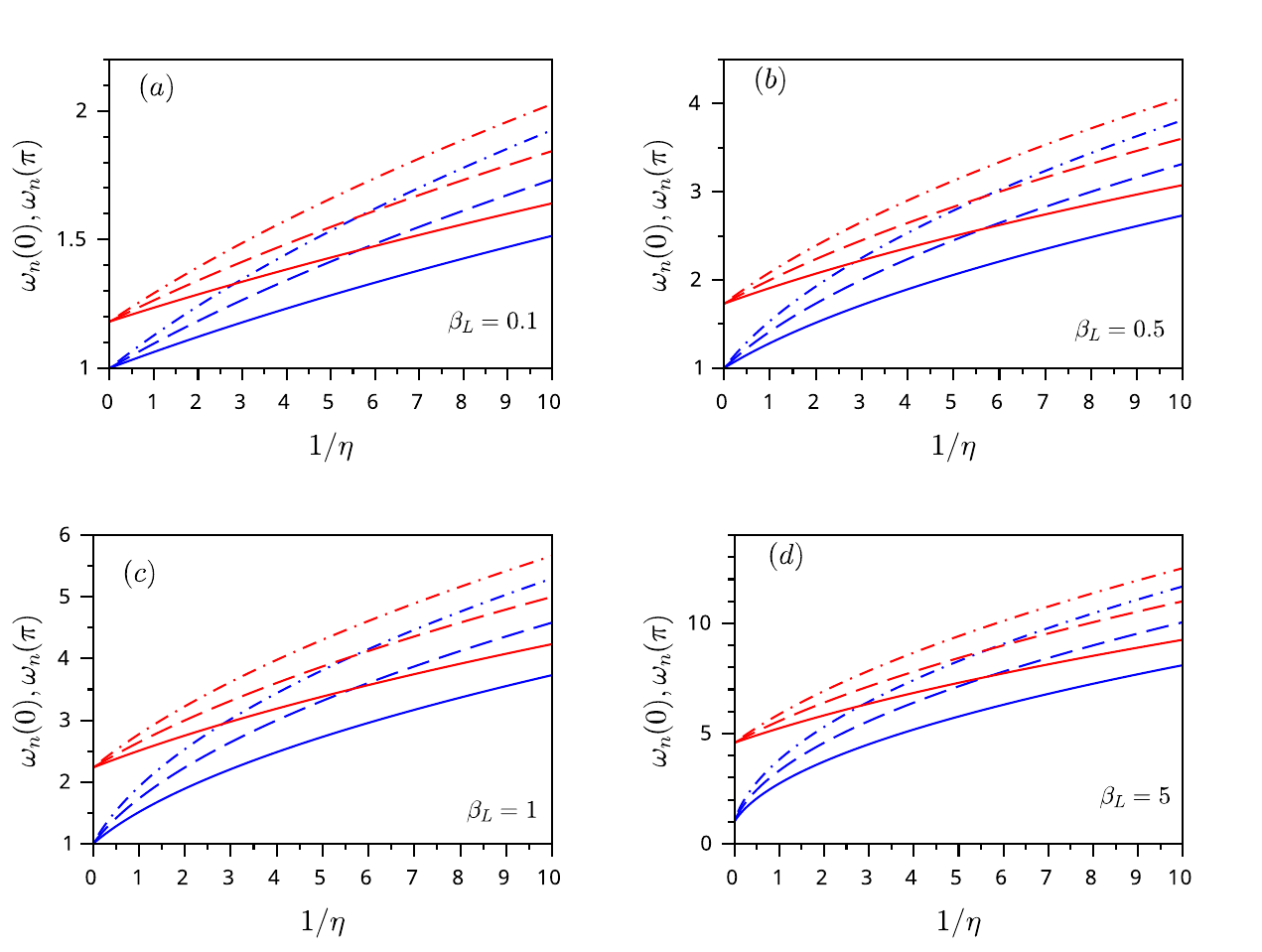}
\caption{Upper (red) and lower (blue) bonds of the strongly dispersive
plasmon branches $\omega_{|n|}$, $n=1,2,3$, for the case of $N=4$ rows with 
the dimensionless bias $\gamma=0.1$ and different values of $\beta_L$. Different type of lines (solid,
$--$ or $-\cdot-$) correspond to a particular branch $\omega_n(q)$.}
\label{bandwidth}
\end{figure}

Before analyzing the plasmon DoS given by Eqs. (\ref{13}) it is useful 
to discuss the behaviour of the edge frequency values $\omega_n(0)$
and $\omega_n(\pi)$ of the strongly dispersive modes ($n>0$).   
They are shown in Fig. \ref{bandwidth} as a function of the inverse 
anisotropy parameter $1/\eta$.
%
%
In the limit $1/\eta \to \infty$ ($\eta \to 0$) the branches are well separated
and there are gaps between them. We remind that this limit corresponds
to the situation when most of the plasma oscillations takes place in
the horizontal subsystem. As $\eta$ increases the gaps start to close
and the dispersion curve begin to overlap.

\begin{figure}[htb]
\includegraphics[scale=0.28]{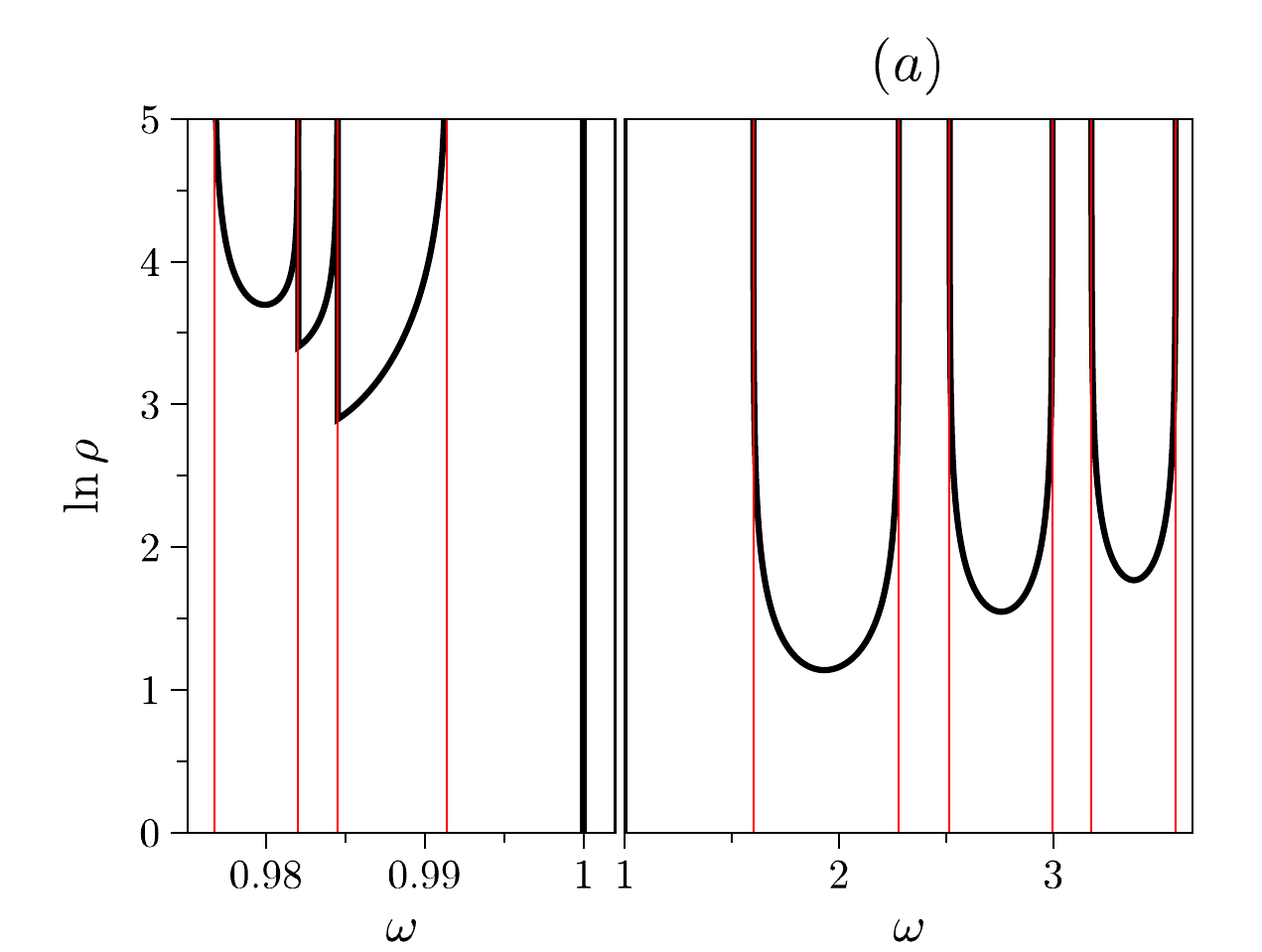}
\includegraphics[scale=0.28]{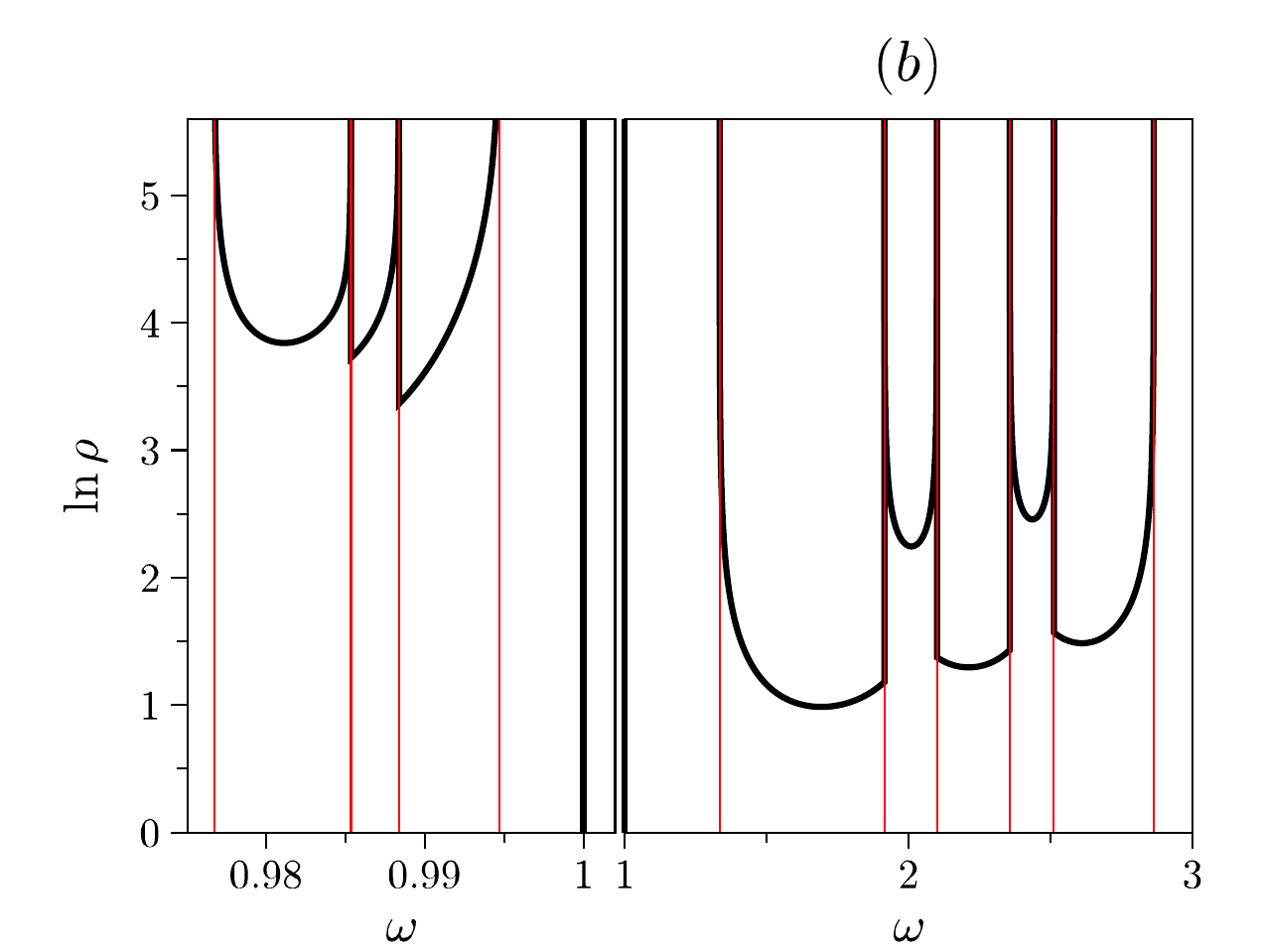}\\
\includegraphics[scale=0.28]{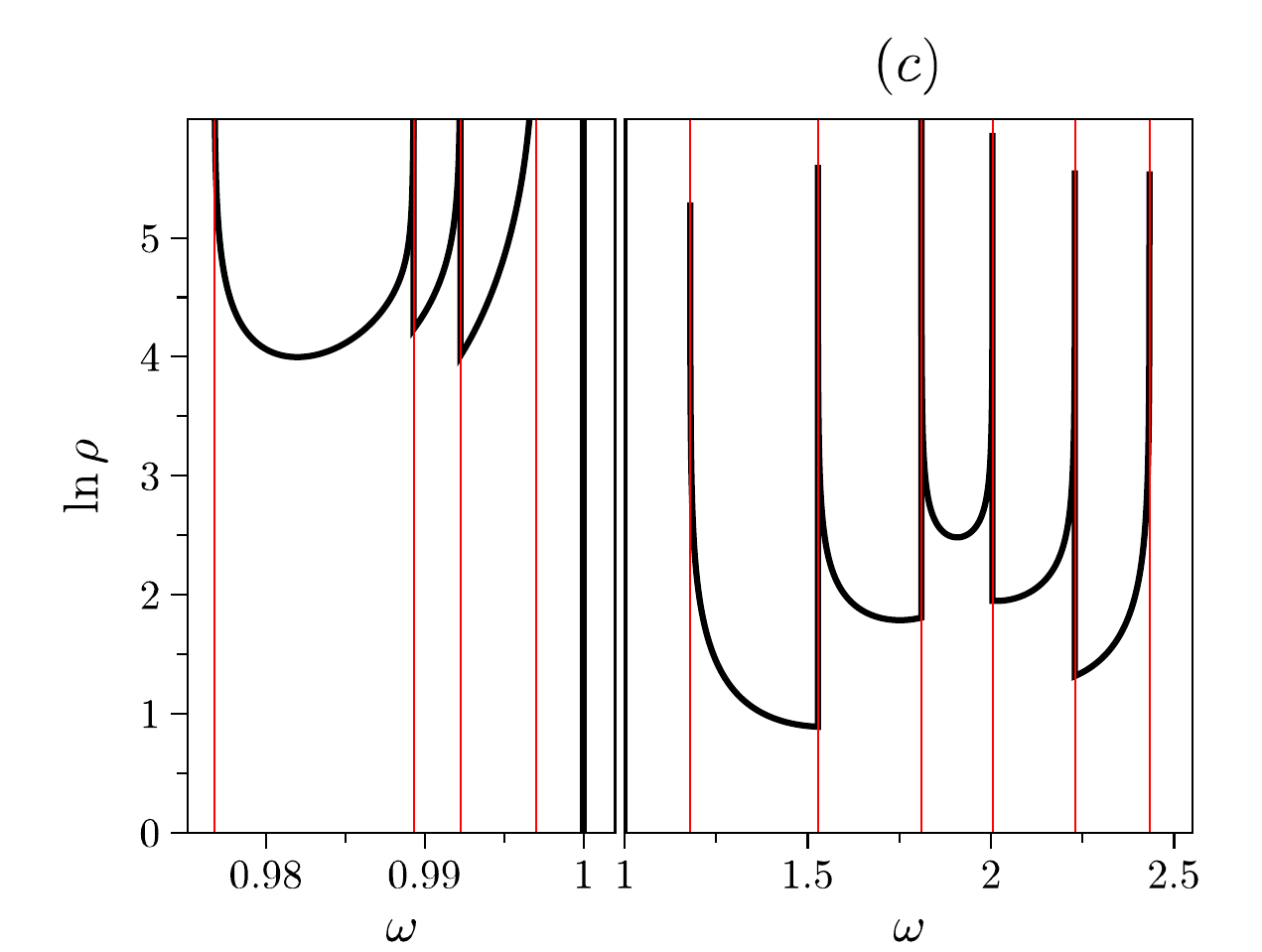}
\includegraphics[scale=0.28]{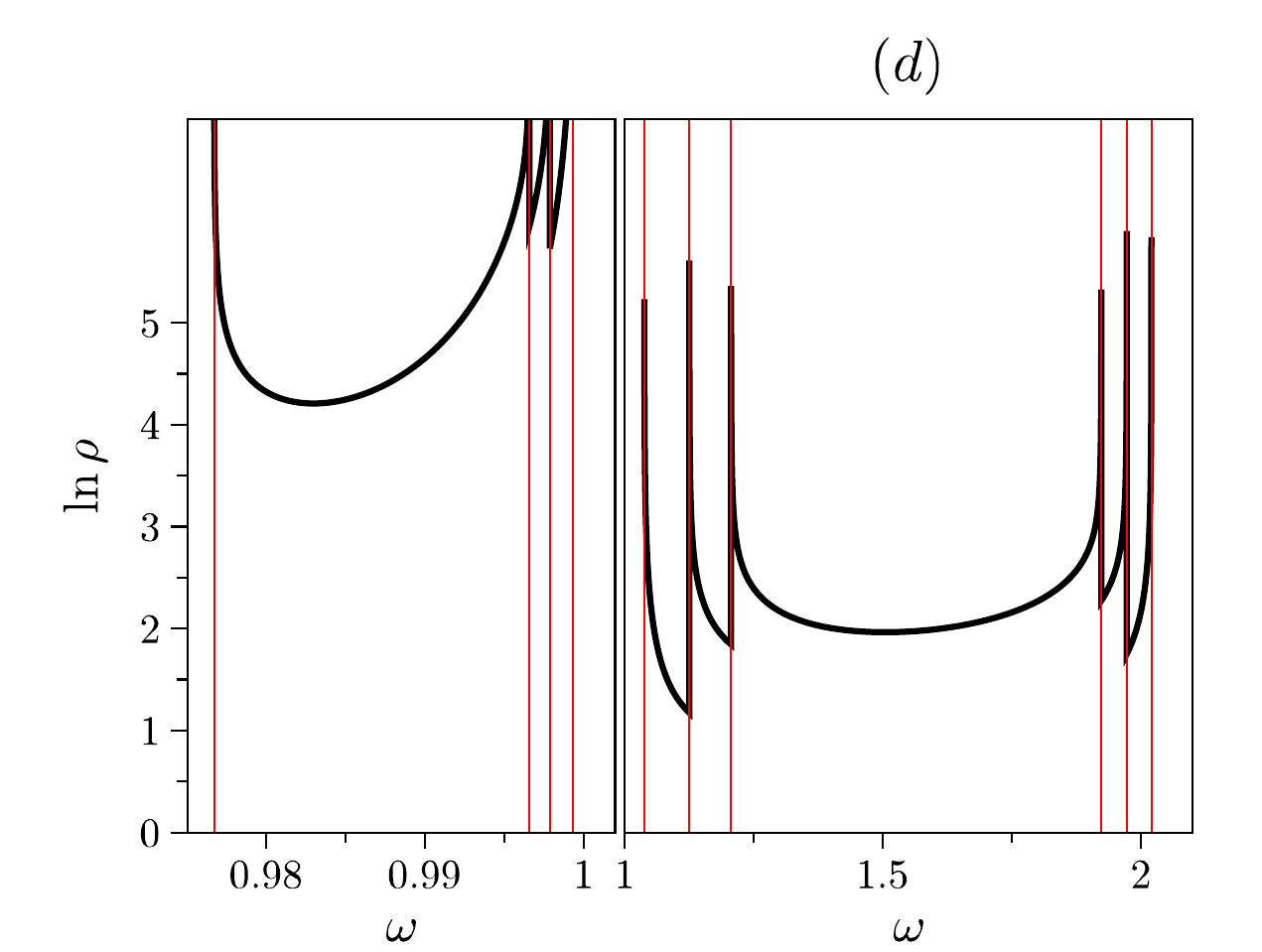}
\caption{Plasmon (log scale) DoS for the array with $N=4$ rows and 
$\gamma=0.3$, $\beta_L=1.5$ and
$\eta=0.25$ (a), $\eta=0.5$ (b), $\eta=1$ (c) and $\eta=5$ (d).
Red vertical lines correspond to edge values $\omega_n(0)$, $\omega_n(\pi)$ 
for each of the plasmon branches.}
\label{dos1}
\end{figure}

\begin{figure}[htb]
\includegraphics[scale=0.37]{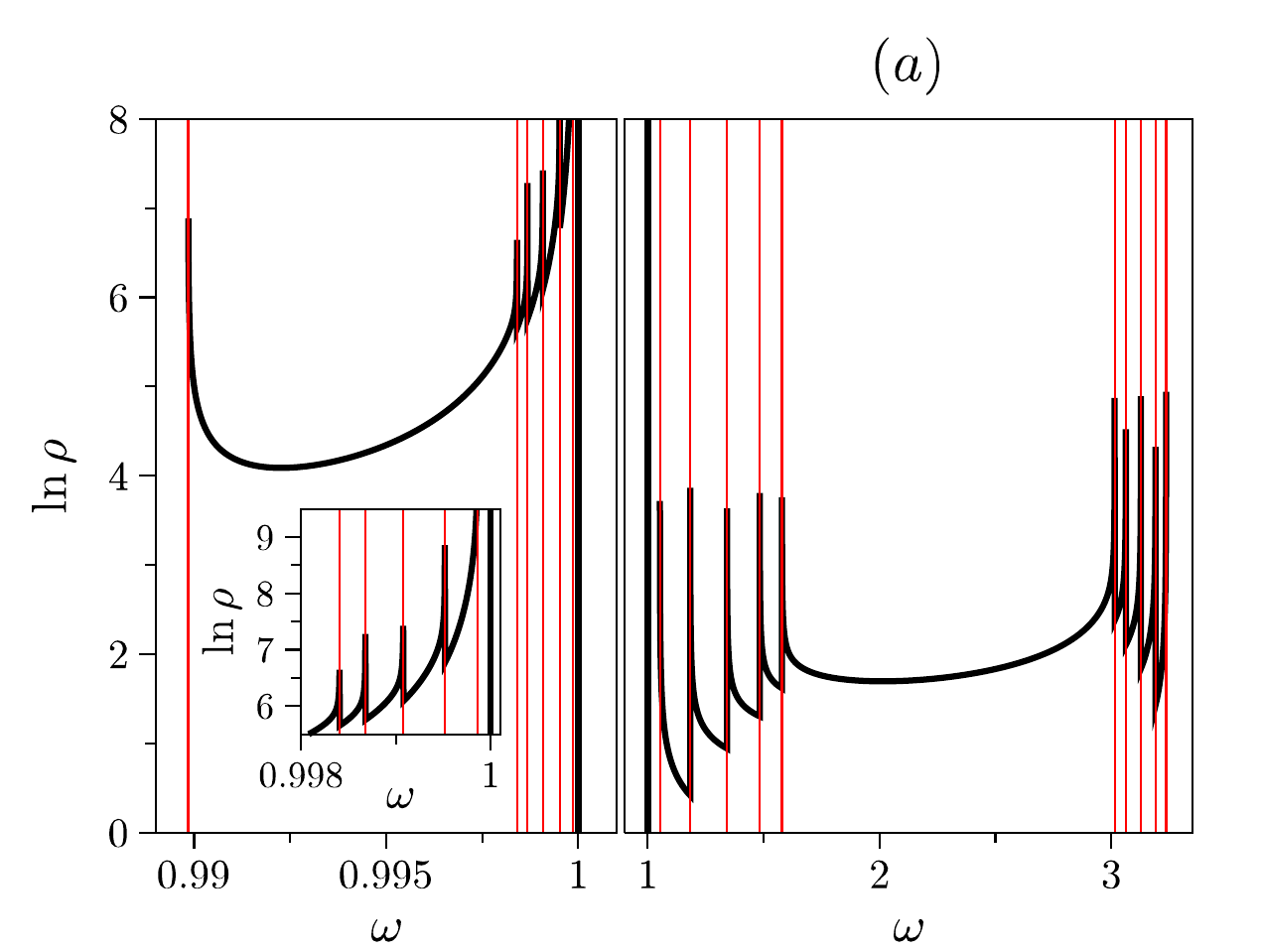}
\includegraphics[scale=0.37]{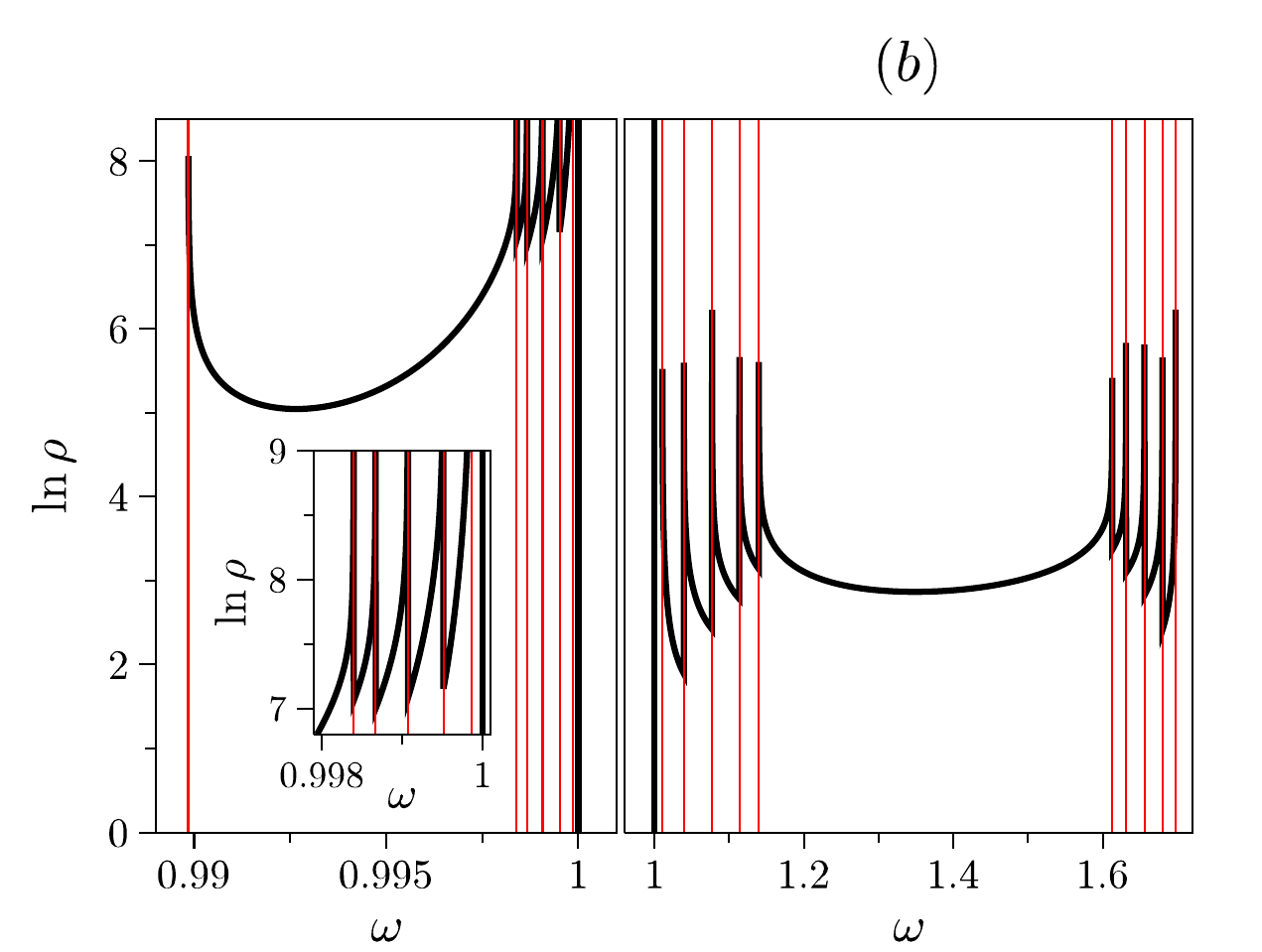}
\caption{Plasmon DoS (log scale) for the array with $N=6$ rows, 
$\gamma=0.2$, $\eta=5$ and
$\beta_L=0.5$ (a) and $\beta_L=2.5$ (b).
Red vertical lines have the same meaning as in the previous figure.
The insets show the detailed picture around the $\omega_{-|n|}(\pi)$
singularities.}
\label{dos2}
\end{figure}

Being equipped with this knowledge we can analyze the plasmon DoS plots
presented in Figs. \ref{dos1}-\ref{dos2}. In those figures the DoS is plotted
as a function of the plasmon frequency for different values of $\eta$.
The situation if Fig. \ref{dos1}(a) corresponds to the non-overlaping
branches.
For $\omega>1$ one can observe three well separated smooth curves with two
van Hove singularities each that correspond to the edge frequency values. As the anisotropy
is increased from $\eta=0.25$ to $\eta=0.5$ the separate branches
begin to overlap [see Fig. \ref{dos1}(b)]. The gaps that were visible
in the previous figure now are closed. The singularities remain but
now $\omega_2(0)<\omega_1(\pi)$ and $\omega_3(0)<\omega_2(\pi)$.  
%
%
Further increasing of $\eta$ causes the singularities at 
$\omega=\omega_{1,2,3}(0)$ to move further to the left, while 
the singularities at $\omega=\omega_{1,2,3}(\pi)$ move to the right.
The limit $\eta \gg 1$ corresponds to the situation when the amplitude
of the oscillations in the vertical subsystem is much larger then
in the horizontal. In this limit the dispersion curves are nested
inside each other rather deeply and the DoS singularities concentrate
at the respective edges of the band [compare Figs. \ref{dos1}(c) and
(d)].

We have not discussed yet the properties of the plasmon DoS in the
frequency region $\omega<1$. At $\omega=1$ we have a perfect 
flat band
and there is a  solid vertical line in Fig. \ref{dos1} to 
demonstrate it. The almost flat branches $\omega_{-|n|}(q)$
are densely squeezed in the interval 
$\omega_{-|n|}(0)\equiv (1-\gamma^2)^{1/4} \le \omega_{-|n|}(q) \le
\omega_{-|n|}(\pi)<1$. Even for rather large bias values this
interval is very narrow. For example, $(1-\gamma^2)^{1/4}\approx 0.931$
for $\gamma=0.5$. The point $q=0$ is $(N-1)$-fold degenerate. 
For the Brillouin zone edge we have $\omega_{-3}<\omega_{-2}<\omega_{-1}$.
The full expressions for $\omega_{-|n|}(\pi)$ is rather complicated and,
therefore, we omit it. So, the dispersion curves are completely
nested inside each other. As a result, we always have $N$ van Hove singularities,
one at $\omega_{-|n|}(0)$ and $N-1$ singularities at $\omega_{-|n|}(\pi)$. When $\eta$
is increased the singularities at $\omega_{-|n|}(\pi)$ concentrate
on the right side of the DoS dependence.
 
In figure \ref{dos2} the dependence of the plasmon DoS on the 
discreteness parameter $\beta_L$ is demonstrated. 
%
%
There are $N=6$ rows, therefore we observe $3N-2=16$ van Hove singularities.
We observe that the change of $\beta_L$ does not bring any 
qualitative difference. The discreteness
parameter influences the total width of the upper ($\omega>1$) 
plasmon band. If $\beta_L$ increases, the discreteness of the array
increases and band narrows; it widens otherwise. Since the width of the
lower ($\omega<1$, almost flat) subband depends mainly on the bias current, 
no visible changes are seen there when $\beta_L$ is varied.
As a final remark we note that the minimal value of DoS is several
orders of magnitude larger for the almost flat branches as compared
to the strongly dispersive branches.

\subsection{Classification of the plasmon modes}

The spatial structure of the plasmon modes that correspond to
each of the branches $\omega_n(q)$ is given by the matrix equation
\begin{equation}\label{9}
\left ( \begin{array}{cc}
[\omega_i^2(q) - \omega^2] \hat{\mathbb{I}}_{N-1} & \frac{1-e^{-iq}}{\beta_L}
\hat{S} \\
\frac{1-e^{iq}}{\beta_L\eta}S^T & \hat{D}_h-\omega^2 \hat{\mathbb{I}}_N  \end{array} \right ) 
\left(\begin{array}{c}
 \mathbf{A}^{(v)} \\ \mathbf{A}^{(h)}
\end{array}\right )=0,
\end{equation}
that appears directly from the equations of motion \cite{bz22jpc}.
Here the vectors $\mathbf{A}^{(v,h)}$ are amplitudes of the plasmon
waves, $\mathbb{I}_N$ is $N\times N$ unit matrix, 
$\hat{D}_h$ is a tridiagonal $N\times N$ matrix 
\begin{eqnarray}\!\!\!\!\!\!\!\!\!\!\!\!\!\!\!\!\!\!\!\!
&&\hat{D}_h=\frac{1}{\eta\beta_L}\left[\begin{array}{ccccccc}\vspace{2mm}
1+\eta\beta_L&-1&0&\cdots &0&0&0\\ \vspace{2mm}
-1&2+\eta\beta_L&-1&\cdots &0&0&0\\ \vspace{2mm}
\vdots &\vdots &\vdots &\ddots & \vdots&\vdots &\vdots\\ \vspace{2mm}
0&0&0&\cdots &-1&2+\eta\beta_L&-1\\ \vspace{2mm}
0&0&0&\cdots &0&-1&1+\eta\beta_L\\
\end{array}\right],
\label{10}
\end{eqnarray}
and $\hat{S}$ is a $(N-1)\times N$ bidiagonal matrix with 
$S_{11}=S_{22}=\cdots=S_{N-1,N-1}=1$, $S_{12}=S_{23}=\cdots=S_{N-1,N}=-1$
and $S_{ij}=0$ elsewhere. We can introduce a normalized real 
$(2N-1)$-component eigenvector
\begin{equation}\label{14}
\mathbf{e}_n(q)=\left( \begin{array}{c}
 \mathbf{e}_n^{(v)} \\ \mathbf{e}_n^{(h)}
\end{array}\right),
\end{equation}
that consists of $N-1$ components for the vertical junctions and
$N$ components for the horizontal junctions.

In the long-wave limit ($q=0$) the horizontal and vertical subsystems
decouple and the respective eigenvectors are easy to find. Previously
only qualitative description of the eigenvectors \cite{bz22jpc} has
been given. In this paper the complete set of solutions will be presented.

\paragraph*{The flat branch $\omega_0=1$} 
For all values of the wavenumber $q$ only horizontal subsystem is excited
while all vertical junctions do not oscillate. 
It can be easily seen from the separate equations for
each of the subsystems: 
$[\omega^2_i(q)-1] \mathbb{I}_{N-1}\mathbf{A}^{(v)}=0$
for the vertical one and $\hat D_h \mathbf{A}^{(v)}=0$ for
the horizontal one. 
Since usually flat bands
appear due to destructive inteference, it easy to understand how such
a dispersionless mode appears in a ladder: when wave propagates
in the $X$ direction, all excitations in the $Y$ direction cancel each other.
The respective normalized eigenvector $\mathbf{e}_0$ equals
\begin{equation}
\mathbf{e}_0=\frac{1}{\sqrt{N}}\left(\underbrace{0, 0, \cdots, 0}_{=\mathbf{e}_0^{(v)}}
\underbrace{ 1, 1, \cdots, 1 }_{=\mathbf{e}_0^{(h)}}\right)^T.
\end{equation}
It can easily be checked that the $\mathbf{A}^{(v)}=0$,
$A^{(h)}_1=\cdots=A^{(h)}_N$ state solves Eq. (\ref{9})
for any $q$ and $\gamma$ if $\omega=1$.

\paragraph*{The long-wave limit $q=0$ for the $\omega_{n}=1$ modes at $\gamma=0$} Note that for $\gamma=0$ the $\omega=1$ branch is
$N$-fold degenerate.
In the vertical subsystem we have 
$0 \cdot \mathbb{I}_{N-1}\mathbf{A}^{(v)}=0$. 
The horizontal subsystem the equality 
$A^{(h)}_1=A^{(h)}_1=\cdots=A^{(h)}_N$ still holds. 
Thus, we can create the normalized eigenvectors 
that solve Eq. (\ref{9}) and are orthogonal to $\mathbf{e}_0$
and each other:
\begin{eqnarray}\!\!\!\!\!\!\!\!\!\!\!\!\ \label{16}
&&\mathbf{e}_{-1}^{(v)}=\left( \begin{array}{c}
1 \\0\\  \cdots \\ 0 \end{array}\right),
\mathbf{e}_{-2}^{(v)}=\left( \begin{array}{c}
0 \\ 1 \\ \cdots \\ 0 \end{array}\right),\ldots,
\mathbf{e}_{-N+1}^{(v)}=\left( \begin{array}{c}
0 \\ 0 \\ \cdots \\ 1 \end{array}\right); \\
&&\mathbf{e}_{-n}^{(h)}=\left( \begin{array}{c}
0 \\ 0 \\ \cdots \\ 0 \end{array}\right),\;\; n=\overline{1,N-1}.
\end{eqnarray}

\paragraph*{Almost flat bands $\omega_{-|n|}$, $n=\overline{1,N-1}$ for $q=0$ and $\gamma\neq 0$}
In this case $\omega_{-|n|}(0)=(1-\gamma^2)^{1/4}\neq 1$. This point
is $(N-1)$-fold degenerate. The vertical subsystem is 
governed by the $0 \cdot \mathbb{I}_{N-1}\mathbf{A}^{(v)}=0$
equation. Thus, the $\mathbf{e}_{-|n|}^{(v)}$
components of the respective eigenvector can be chosen the
same as in Eq. (\ref{16}). One of the options is to keep 
components of the horizontal
subsystem $\mathbf{A}^h=0$ since it is a solution. 
There is no other nontrivial solution (see \ref{app1} for details).
Thus, we need to keep the same solutions as in the previous
subparagraph.

\paragraph{Strongly dispersive branches $\omega_{n>0}$} The susbsystems
stay decoupled, the vertical subsystem is not exited while
in the horizontal system the plasmon amplitudes are non-zero. The 
components of the eigenvector are expressed through
the polynomials ${\cal T}_k(\xi)$ with the
recurrence relation ${\cal T}_{k+1}(\xi)=(\xi+1){\cal T}_{k+1}(\xi)-{\cal T}_{k-1}(\xi)$, which are defined in \ref{app1}:
\begin{eqnarray}\label{28}
\mathbf{e}_n^{(h)}=\frac{1}{\sqrt{\cal N}_n}\left [ \begin{array}{c}
1 \\ {\cal T}_2(-\alpha_{n+1}) \\ {\cal T}_3(-\alpha_{n+1}) \\
\cdots \\ {\cal T}_{N-1}(-\alpha_{n+1}) \\ {\cal T}_{N}(-\alpha_{n+1}) \end{array} \right ],\;\;
{\cal N}_n=\sum_{k=1}^{N}{\cal T}^2_{k}(-\alpha_{n+1}).
\end{eqnarray}
There is also the boundary condition 
${\cal T}_{N-1}(-\alpha_{n+1})=-\alpha_{n+1}{\cal T}_{N1}(-\alpha_{n+1})$
which should be satisfied. In fact, it satisfies automatically.
Note that sometimes $\alpha_{n+1}=0$. This happens, for example, 
for the $N=3$ row array,
for the mode $\omega_1(0)$. In that case $\alpha_2=0$. 
In general, this will happen for arrays with $N=3m$, $m=1,2,3,\dots$,
because there will always exist some mode $n$ with $\cos \pi/3=1/2$
in the coefficient $\alpha_{n+1}$. In those cases the recurrence relation together
with the boundary conditions will produce the following normalized
eigenvector:
\begin{eqnarray}\label{29}
\mathbf{e}_n^{(h)}=\frac{1}{\sqrt{2N/3}}
\left [1,0,-1,-1,0,1,1,0,-1,-1,0,1,\ldots,0,(-1)^{N/3}  \right ]^T.
\end{eqnarray} 
The power $(-1)^{N/3}$ in the last component can be easily explained
by the fact that these eigenvectors consist of alternating 
triples $(1,0,-1)$ and depending on whether $N/3$ is odd or even
the eigenvector will end with $-1$ or $1$.

Let us produce some examples.
For the array with $N=3$ rows there are two strongly
dispersive modes $\omega_1$ and $\omega_2$. There are two respective
normalized eigenvectors:
\begin{equation}\label{26}
\!\!\!\!\!\!\!\!\!\!\!\!\!\!\!
\mathbf{e}_{n}^{(v)}=\left( \begin{array}{c}
0 \\0 \end{array}\right),
\mathbf{e}_{1}^{(h)}=\frac{1}{\sqrt{2}}\left( \begin{array}{c}
1 \\ 0 \\ -1 \end{array}\right),
\mathbf{e}_2^{(h)}=\frac{1}{\sqrt{6}}\left( \begin{array}{c}
1 \\ -2 \\ 1 \end{array}\right).
\end{equation}
The eigenvector for $\omega_1$ has $\alpha_2=0$, thus its structure
is defined by (\ref{29}).
Similarly, for the $N=4$ rows there are three highly dispersive modes $\omega_{1,2,3}$ and the respective the eigenvectors read
\begin{eqnarray}\label{27}
\!\!\!\!\!\!\!\!\!\!\!\!\!\!\!\!\!\!\!\!\!\!\!\!\!\!\!\!\!\!
&&\mathbf{e}_{n}^{(v)}=\left( \begin{array}{c}
0 \\0 \\0 \end{array}\right),
\mathbf{e}_{1}^{(h)}=\frac{1}{2\sqrt{2-\sqrt{2}}}\left( \begin{array}{c}
1 \\ \sqrt{2}-1 \\ 1-\sqrt{2} \\ -1 \end{array}\right),
\mathbf{e}_2^{(h)}=\frac{1}{\sqrt{2}}\left( \begin{array}{c}
1 \\ -1 \\ -1 \\1 \end{array}\right),\\
\!\!\!\!\!\!\!\!\!\!\!\!\!\!\!\!\!\!\!\!\!\!\!\!\!\!\!\!\!\!
&&\mathbf{e}_3^{(h)}=\frac{1}{\sqrt{7+2\sqrt{2}}} \left( \begin{array}{c}
1 \\ -1-\sqrt{2} \\ 1+\sqrt{2} \\-1 \end{array}\right).
\end{eqnarray}
When these component are calculated the boundary condition for the last
two components is satisfied automatically. The reader is welcome to check it.

For the $N=6$ rows we have $5$ strongly dispersive branches,
the coefficients $\alpha_{n+1}$ are: $\alpha_2=1-\sqrt{3}$,
$\alpha_3=0$, $\alpha_4=1$, $\alpha_5=2$ and $\alpha_6=1+\sqrt{3}$.
The eigenvector for the $n=2$ branch ($\alpha_3=0$) can be easily
computed from the recurrence relations: $A_1=1$, $A_2=0$,
$A_3=A_2-A_1=-1$, $A_4=A_3-A_2=-1$, $A_5=0$, $A_6=A_5-A_4=1$.
Thus, we have $\mathbf{e}^{(h)}_2=(1/2)(1,0,-1,-1,0,1)^T$ which
satisfies the general Eq. (\ref{29}).

\paragraph*{Non-flat bands for $q\neq 0$} Finally we discuss all
branches $\omega_{\pm n}(q)$, $n=1,2,\ldots,N-1$. 
In this case the phases
from both the horizontal subsystem are involved in the dynamics.
The amplitudes for the horizontal subsystems satisfy exactly
the same relations (\ref{28}) (see  \ref{app1} for details).
In fact, both the strongly dispersive ($\omega_{|n|}$)
and almost flat ($\omega_{-|n|}$) modes have the same distribution 
in of the horizontal oscillatons.
The amplitudes of the vertical subsystem
can be expessed through the same polynomials ${\cal T}_k$
(see \ref{app1} for details):
\begin{eqnarray}\!\!\!\!\!\!\!\!\!\!\!\! \nonumber
&&\mathbf{e}^{(v)}_{\pm n}(q)=\frac{1}{\sqrt{{\cal N}_n(q)}} 
\frac{1-\cos{q}}{\beta_L[\omega^2_{\pm n}(q)-\omega^2_i(q)]} 
\left[\begin{array}{c} 
1+\alpha_{n+1} \\ {\cal T}_2(-\alpha_{n+1})- {\cal T}_3(-\alpha_{n+1}) \\ {\cal T}_3(-\alpha_{n+1})- {\cal T}_4(-\alpha_{n+1})\\ \cdots \\
 {\cal T}_{N-2}(-\alpha_{n+1})-{\cal T}_{N-1}(-\alpha_{n+1}) \\
 {\cal T}_{N-1}(-\alpha_{n+1})-{\cal T}_{N}(-\alpha_{n+1})  \end{array}  \right],\\ \label{33}
\end{eqnarray}
with the normalization coefficient
\begin{eqnarray}\label{29a}
{\cal N}_n(q)&=&\sum_{k=1}^{N} {\cal T}_k^2(-\alpha_{n+1})+ 
\frac{(1-\cos{q})^2}{\beta^2_L[\omega^2_{\pm n}(q)-\omega^2_i(q)]^2}
\times  \\
& \times &  \left \{
\sum_{k=1}^{N-1} \left[ {\cal T}_k(-\alpha_{n+1})-{\cal T}_{k+1}(-\alpha_{n+1})\right]^2 \right\} .
\nonumber
\end{eqnarray}

We see that different modes, $\omega_{n}$ and $\omega_{-n}$ with 
demonstrate the same dynamics in the horizontal subsystem. Thus,
the difference between these two types of modes lie in the
dynamics of the vertical phases. If one focuses on the
coefficient in front of the vector in (\ref{33}) in the limit of 
small $q$'s, one can see that
for the strongly dispersive modes it equals 
\begin{equation}
\frac{1}{2\beta_L \left (1-\sqrt{1-\gamma^2}+\frac{1+\alpha_{n+1}}{\eta\beta_L} \right)} q^2 +{\cal O}(q^4).
\end{equation}
For the almost flat bands ($\omega_{-n}$) the denominator of the same
coefficient is $\propto q^2$. Its expansion has the following form: 
\begin{equation}
\frac{1-\sqrt{1-\gamma^2}}{\sqrt{2}\beta_L\sqrt{1-\gamma^2}
\left( 1- \sqrt{1-\gamma^2} +\frac{1+\alpha_{n+1}}{\eta\beta_L}\right) }
+{\cal O} (q^2). 
\end{equation}
Thus, the vertical subsystem behaves differently for the
two types of modes. The strongly dispersive modes do not have
a constant term, while the almost flat ones do.

From the properties of the ${\cal T}_k$ polynomials (see Fig. \ref{poly}) we can see how the structure of the amplitudes of the vertical and horizontal junctions. They can be in phase, in anti-phase or some
more complicated picture can appear. The special cases of
particular vaues of $-\alpha_{n+1}=0,-2$ can easily
be spotted.  Here are some examples of the amplitude distribution in the vertical subsystem. For the $N=3$ case the $\mathbf{e}_{\pm 1}^{(v)}$
vector can be derived from the  $\mathbf{e}_{\pm 1}^{(h)}$
vector (\ref{26}): $\mathbf{e}_{\pm 1}^{(v)} \propto (1,1)^T$.
In the same way we have another vector $\mathbf{e}_{\pm 2}^{(v)} \propto (1,-1)^T$. So, the vertical subsystem have both the in-phase and
anti-phase modes. As another example, let us consider the $\omega_{\pm 2}$
mode of the $N=6$ array. The horizontal part of the respective
eigenvector was obtained previously: $\mathbf{e}^{(h)}_{\pm 2}=(1/2)(1,0,-1,-1,0,1)^T$. The respective vertical part is
$\mathbf{e}^{(v)}_{\pm 2} \propto (1,1,0,-1,-1)^T$.

\section{Conclusions}

This work is devoted to the spectral properties of the Josephson 
plasma waves (Josephson plasmons) in the inductively coupled
quasi-one-dimensional ladder-like array of Josephson junctions. 
An intriguing feature of this array is a $N-$fold degenerate flat
band in the unbiased case. If an external bias is applied 
the degeneracy is lifted for all wavenumbers
except the $q=0$ point and the set of very weakly dispersive or 
almost flat bands appears.

The plasmon density of states (DoS) has $3N-2$ van Hove singularities that
appear at the edge values of the individual branches and one
$\delta$-function-like peak due to the flat band. There are $N$
singularities for the almost flat subband that lies under the
$\omega=1$ (in the units of the Josephson plasma frequency) branch and
$2(N-1)$ for the strongly dispersive subband that lies above the
$\omega=1$ branch.

The eigenvectors that describe the spatial distribution of the
Josephson phase vibrations within the elementary cell have been
obtained. They are expressed through the set of orthogonal polynomials
that have a recurrence relation similar as in the Chebyshev
polynomials but still different. It should be noted that  
Chebyshev-like polynomials of more general form 
have appeared before for an unrelated system that
also involves nearest-neighbour lattice interations \cite{ze01prb}.

We believe that these results will be important for the studies
of the nonlinear excitations in Josephson ladders.
For example, fluxon interaction with the plasma waves in the
simple parallel array gives a rise to the novel resonant 
effects \cite{ucm93prb}. Since the $N-$row ladder has such a rich
linear spectrum, it would be intriguing to study the fluxon-plasmon
interaction in such a system.

\section*{Acknowledgements}
We would like to thank the Armed Forces of Ukraine for providing security
to perform this work.
I.O.S. and Y.Z. acknowledge support from the National
Research Foundation of Ukraine, grant (2023.03.0097)
"Electronic and transport properties of Dirac materials and
Josephson junctions". 

\appendix
\section{Details of the eigenvector calculation}
\label{app1}
In this Appendix the procedure of the plasmon amplitude calculation is presented.
The plasmon amplitudes $(\mathbf{A}^{(v)},\mathbf{A}^{(h)})$ 
do not depend on the row number $m$, but 
depend on the column number. They are governed by the
set of matrix equations (\ref{9}) which can be rewritten as
\begin{eqnarray}\label{A1}
&& [\omega_i^2(q) - \omega^2] {\hat \mathbb{I}}_{N-1}\mathbf{A}^{(v)}+
\frac{1-e^{-iq}}{\beta_L}\hat{S} \mathbf{A}^{(h)}=0\\
&& \frac{1-e^{iq}}{\beta_L\eta}\hat{S}^T\mathbf{A}^{(v)}+(\hat{D}_h-\omega^2 {\hat\mathbb{I}}_N)\mathbf{A}^{(h)}=0.\label{A2}
\end{eqnarray}
Here $\omega$ is the plasmon frequency that belongs to
any of the branches (\ref{7})-(\ref{8}).
After substituting the explicit form of the matrices $\hat{D}_h$,
$\hat{S}$ and $\hat{S}^T$ we rewrite the equations (\ref{A1})-(\ref{A2}).
The first equation allows us to express the vertical amplitudes
through the horizontal ones:
\begin{eqnarray}\label{A3}
[\omega_i^2(q) - \omega_n^2]A_k^{(v)}+\frac{1-e^{-iq}}{\beta_L} 
 \left( A_k^{(h)}-A_{k+1}^{(h)}\right)=0,\; k=\overline{1,N-1}.
\end{eqnarray} 
The second equation takes the following form
\begin{eqnarray}
\label{A4}
&& ({1-e^{iq}}) \left (A_k^{(v)}-A_{k-1}^{(v)}\right )- A_{k-1}^{(h)}- A^{(h)}_{k+1}+\\
\nonumber 
&&+[\beta_L\eta(1-\omega^2)+2]A_k^{(h)}=0,\;k=\overline{2,N-1},\\
\label{A5}
&&({1-e^{iq}}) A_1^{(v)}+ [\beta_L\eta(1-\omega^2)+1]
A_1^{(h)}-A_2^{(h)}=0,\\
\label{A6}
&&({1-e^{iq}}) A_N^{(v)} +A_{N-1}^{(h)}-[\beta_L\eta(1-\omega^2)+1]
A_N^{(h)}=0 ~.
\end{eqnarray}

In the long wave limit $q=0$ the prefactors $1-e^{\pm iq}$ in Eqs.
(\ref{A1})-(\ref{A2}) vanish and, consequently, the vertical and 
horizontal subsystems are decoupled:
\begin{equation}\label{A7}
[\sqrt{1-\gamma^2} - \omega^2] \hat{\mathbb{I}}_{N-1}\mathbf{A}^{(v)}=0,
\;\; [ \hat{D}_h-\omega^2 \hat{\mathbb{I}}_N] \mathbf{A}^{(h)}=0.
\end{equation}
The strongly degenerate case of $q=0$, $\gamma=0$ has been discussed in the main text.
 
\begin{figure}[htb]
\includegraphics[scale=0.25]{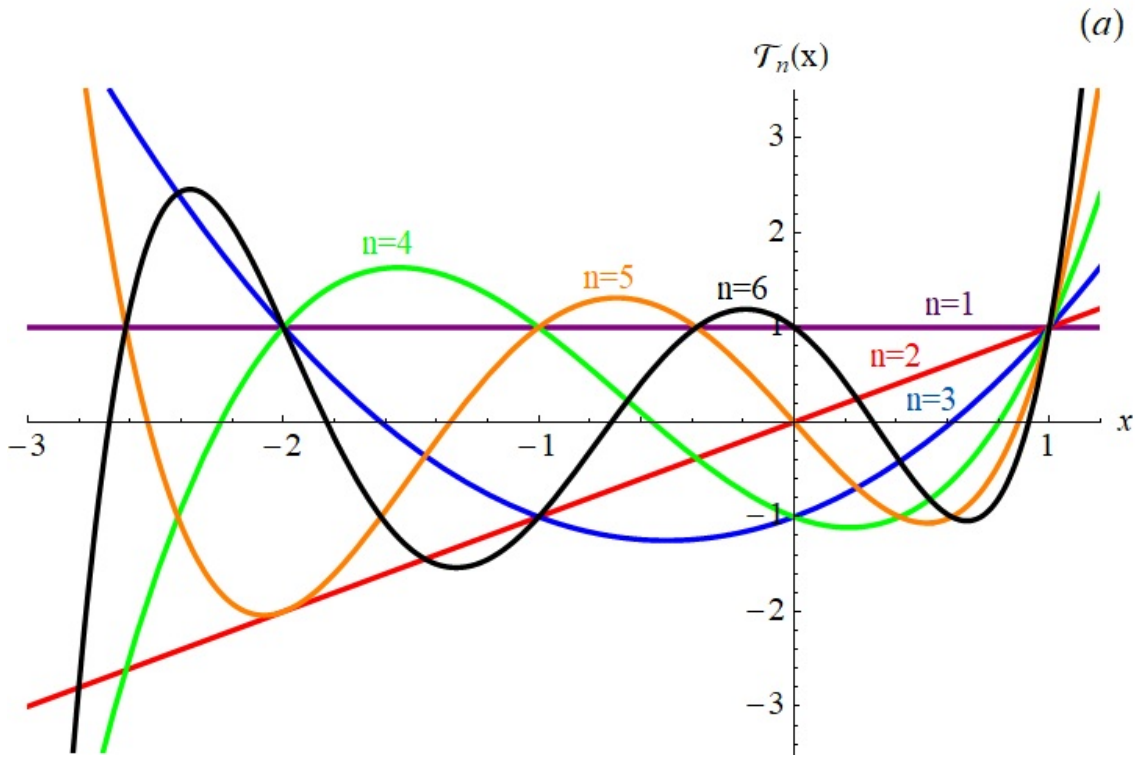}
\includegraphics[scale=0.25]{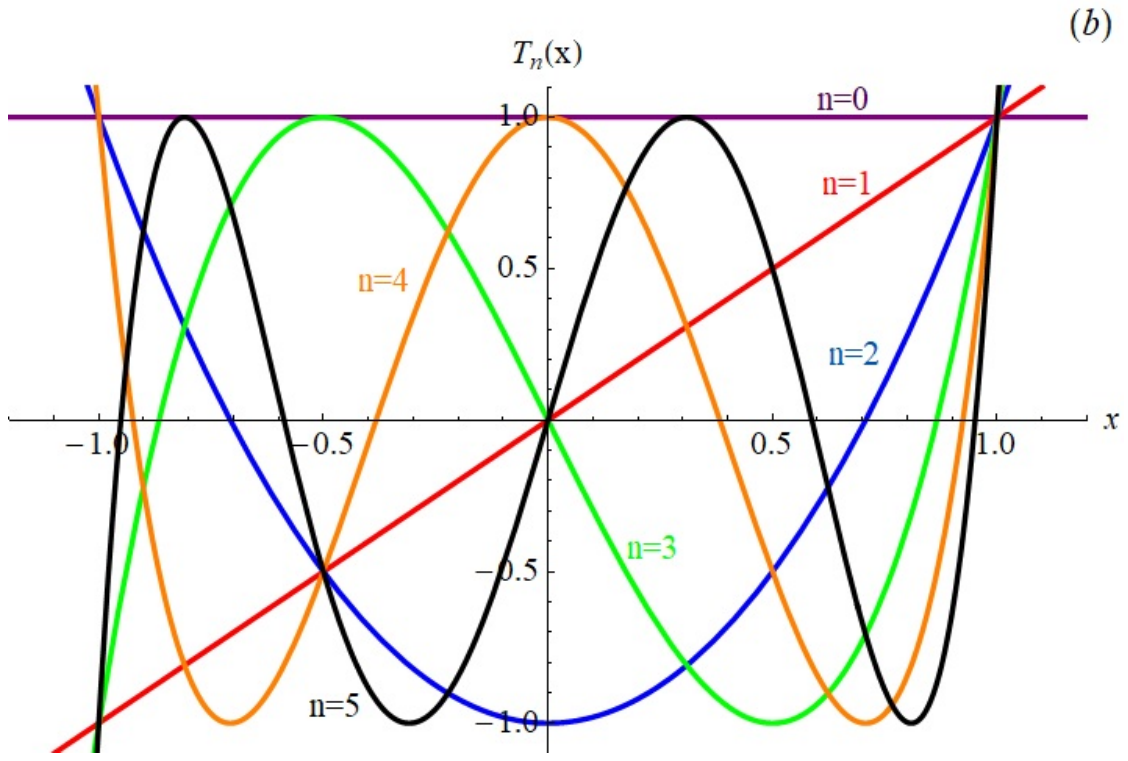}
\caption{Example of several polynomials ${\cal T}_k$ (a) and
Chebyshev polynomials (b) of the first kind.}
\label{poly}
\end{figure}

For the dispersive modes $\omega_n(0)\neq 1$, $n\neq 0$, the amplitudes in the
horizontal subsystem generally satisfy the system
\begin{eqnarray}\label{A8}
&&A_2^{(h)}=[\beta_L\eta(1-\omega^2)+1] A_1^{(h)},\\
&& \label{A9}
A_N^{(h)}=[\beta_L\eta(1-\omega^2)+1]^{-1} A_{N-1}^{(h)},\\
&& A_{k+1}^{(h)}-[\beta_L\eta(1-\omega^2)+2] A_{k}^{(h)}+
A_{k-1}^{(h)}=0,\; k=\overline{2,N-1}. \label{A10}
\end{eqnarray}
The solutions of this system are real.
It appears that these horizontal amplitudes satisfy the recurrence
relation (\ref{A10}). Without loss of generality we can assume
$A_1^{(h)}=1$, because if $A_1^{(h)}=0$ then all other amplitudes
must be zero.  
Then the components from $k=2$ through $k=N-1$
can be expressed via the following polynomials:
\begin{eqnarray}
&&{\cal T}_1(\xi)=1, \label{A11}\\
&&{\cal T}_2(\xi)=\xi,\\
&&{\cal T}_3(\xi)=(1+\xi){\cal T}_2(\xi)-{\cal T}_1(\xi)=\xi^2+\xi-1,\\
&&{\cal T}_4(\xi)=(1+\xi){\cal T}_3(\xi)-{\cal T}_2(\xi)=\xi^3+2\xi^2-\xi-1,\\
&& \;\;\; \cdots \cdots \cdots \nonumber \\
&& {\cal T}_{k+1}(\xi)=(1+\xi){\cal T}_k(\xi)-{\cal T}_{k-1}(\xi).
\label{A15}
\end{eqnarray}
These polynomials look similar to the well-known Chebyshev polynomials
\cite{as84} but their recurrence relation is different. 
See the respective graphs in Fig. \ref{poly}.
%
%
%

For the case of almost flat branches, $\omega_{-|n|}$, 
$q=0$, $\gamma\neq 0$ the horizontal subsystem can be rewritten through
these polynomials as functions of $\zeta=1+\beta_L\eta(1-\sqrt{1-\gamma^2})$:
${\cal T}_1(\zeta)=1$, ${\cal T}_2(\zeta)=\zeta$ and so on.
Since $\zeta \neq 0$, the last component must satisfy the boundary condition
\begin{equation}\label{A12}
\zeta {\cal T}_N(\zeta)={\cal T}_{N-1}(\zeta).
\end{equation}
At the first glance, we have
a nontrivial solution for this case. However, it is not true
because it is impossible to satisfy (\ref{A12}) for $\zeta>1$.
Indeed, for $\zeta>1$ all the polynomials ${\cal T}_{k}$ for $\zeta>1$
are monotonically increasing functions (see Fig. \ref{poly}). The leading
term of the ${\cal T}_{k}(\zeta)$ polynomial for $\zeta>1$ is 
$\zeta^{k-1}$. There is no way that two functions, ${\cal T}_{k}(\zeta)$
and $\zeta {\cal T}_{k+1}(\zeta)$ would cross anywhere for $\zeta>1$.
 
In the case of strongly dispersive branches $\omega_{|n|}(0)$ we still
have to express the solution of the horizontal subsystem (\ref{A8})-(\ref{A10}) through the same polynomials but
arguments of the polynomials will be different.
According to (\ref{8}) the plasmon frequency for the strongly
dispersive mode is
\begin{equation}\label{A15a}
\omega_n(0)=\sqrt{1+\frac{1}{\eta \beta_L}(1+\alpha_{n+1})},\;\; n=\overline{1,N-1}.
\end{equation}
We substitute it instead of $\omega$ in Eqs. (\ref{A8})-(\ref{A10}).
As a result, we have the polynomials as functions of
\begin{equation}\label{A15b}
\xi_n=\beta_L\eta[1-\omega_n^2(0)]+1=-\alpha_{n+1}=-1+2\cos \frac{\pi n}{N}.
\end{equation}
In some cases we may have $\alpha_n=0$. Then $A^{(h)}_{N-1}=0$ due to use the boundary condition (\ref{A9}),
$A^{(h)}_2=0$, $A^{(h)}_3=-A^{(h)}_2=-1$ and so on according to
the recurrence (\ref{A15}): ${\cal T}_{k+1}(0)={\cal T}_k(0)-{\cal T}_{k-1}(0)$. The components will form the following sequence:
$(1,0,-1,-1,0,1,1,0,-1,-1,0,1,1\ldots)$. This can happen only
if $n/N=1/3$ in (\ref{A15b}), thus the dimension of the vector 
$\mathbf{A}^{(h)}$ should be divisible by $3$. In that case the
last component would be $+1$ or $-1$. 
Hence, they can be written in the followin general form:
\begin{equation}\label{A16}
\mathbf{A}^{(h)}_n(q)=
\left[\begin{array}{c}
{\cal T}_1(-\alpha_{n+1}) \\ {\cal T}_2 (-\alpha_{n+1}) \\ 
{\cal T}_3(-\alpha_{n+1}) \\ \cdots \\
{\cal T}_{N-1}(-\alpha_{n+1}) \\ {{\cal T}_{N}(-\alpha_{n+1})
} \end{array}  \right],\; n=\overline{1,N-1}.
\end{equation}
It can be shown that the boundary condition  (\ref{A9}) is satisfied
automatically.

Finally, consider the the general case $q\neq 0$ for all modes 
$\omega_{\pm n}(q)$, $n=\overline{1,N-1}$. 
We substitute $A^{(v)}_k$ 
from Eq. (\ref{A3}) into Eqs. (\ref{A4})-(\ref{A6}). As a result
we get a pair of equations for the top and bottom of the array
\begin{eqnarray}
A^{(h)}_2= \xi_{\pm n} A^{(h)}_1, \; \xi_{\pm n} A^{(h)}_N=A^{(h)}_{N-1}.
\end{eqnarray}
Similarly, for the rest of the primitive cell  ($k\neq 1$ or $k\neq N$):
\begin{equation}
A_{k-1}^{(h)}+A_{k+1}^{(h)} - ( 1+\xi_{\pm n})A_{k}^{(h)}=0,
\end{equation}
where
\begin{equation}\label{A18}
\xi_{\pm n}=1+
\eta \beta_L\frac{(1-\omega_{\pm n}^2)(\omega_{\pm n}^2-\omega_i^2)}{
\omega_{\pm n}^2-\sqrt{1-\gamma^2}}.
\end{equation}
Since the dispersion law satisfies the biquadratic equation
\begin{equation}
(\omega_{\pm n}^2-\omega_i^2)(\omega_{\pm n}^2-1)-\frac{\alpha_{n+1}+1}{\eta\beta_L}(\omega_{\pm n}^2-\omega_i^2)-\frac{2(1+\alpha_{n+1})(1-\cos q)}{\eta\beta^2_L}=0,
\end{equation}
it is not difficult to figure out that the parameter $\xi_{\pm n}$ in (\ref{A18})
does not depend on $q$, moreover, it equals:
\begin{equation}
\xi_{\pm n}=-\alpha_{n+1}=-1+2\cos \frac{\pi n}{N},
\end{equation}
exactly in the same way as in the $q=0$ case.
Thus, the horizontal components of the plasmon wave  coincide with the $q=0$ case (\ref{A16}). 

The eigenvector components $\mathbf{A}^{(v)}$ 
for the vertical subsystem can be easily written
with the help of Eq. (\ref{A3}). In order to have only the real
part of the amplitude we compute the following quantity:
\begin{eqnarray}\nonumber
&&\frac{1}{2}[A^{(v)}+c.c.]=\frac{1-\cos{q}}{\beta_L[\omega^2_{\pm n}(q)-\omega^2_i(q)]}
\left (\begin{array}{c} A^{(h)}_1-A^{(h)}_2 \\  A^{(h)}_2-A^{(h)}_3 \\
\cdots \\ A^{(h)}_{N-2}-A^{(h)}_{N-1} \\ A^{(h)}_{N-1}-A^{(h)}_N \end{array}\right )=\\
&&=
\frac{1-\cos{q}}{\beta_L[\omega^2_{\pm n}(q)-\omega^2_i(q)]} \left[\begin{array}{c} 
1+\alpha_{n+1} \\ -\alpha_{n+1}- {\cal T}_3(-\alpha_{n+1}) \\ {\cal T}_3(-\alpha_{n+1})- {\cal T}_4(-\alpha_{n+1})\\ \cdots \\
 {\cal T}_{N-2}(-\alpha_{n+1})-{\cal T}_{N-1}(-\alpha_{n+1}) \\
 {\cal T}_{N-1}(-\alpha_{n+1})-{\cal T}_{N}(-\alpha_{n+1})  \end{array} \right ].
\label{A20}
\end{eqnarray}

As a final remark, we would like to note that the polynomials
(\ref{A11})-(\ref{A15}) are orthogonal due to the Farvard's theorem
\cite{favard35}.


\end{document}